# arXiv e-prints and the journal of record:
# An analysis of roles and relationships

Vincent Larivière[1], Cassidy R. Sugimoto[2], Benoit Macaluso[1], Staša Milojević[2], Blaise Cronin[2], and Mike Thelwall[3]

[1] *vincent.lariviere@umontreal.ca*; *macaluso.benoit@uqam.ca*
École de bibliothéconomie et des sciences de l'information, Université de Montréal, C.P. 6128, Succ. Centre-Ville, Montréal, QC. H3C 3J7 (Canada) and
Observatoire des sciences et des technologies (OST), Centre interuniversitaire de recherche sur la science et la technologie (CIRST), Université du Québec à Montréal, CP 8888, Succ. Centre-Ville, Montréal, QC. H3C 3P8, (Canada)

[2] *sugimoto@indiana.edu*; *smilojev@indiana.edu*; *bcronin@indiana.edu*
School of Information and Library Science, Indiana University Bloomington
1320 E. 10th St. Bloomington, IN 47401 (USA)

[3] *m.thelwall@wlv.ac.uk*
School of Technology, University of Wolverhampton, Wulfruna Street, Wolverhampton WV1 1LY (UK).

**Abstract**

Since its creation in 1991, arXiv has become central to the diffusion of research in a number of fields. Combining data from the entirety of arXiv and the Web of Science (WoS), this paper investigates (a) the proportion of papers across all disciplines that are on arXiv and the proportion of arXiv papers that are in the WoS, (b) elapsed time between arXiv submission and journal publication, and (c) the aging characteristics and scientific impact of arXiv e-prints and their published version. It shows that the proportion of WoS papers found on arXiv varies across the specialties of physics and mathematics, and that only a few specialties make extensive use of the repository. Elapsed time between arXiv submission and journal publication has shortened but remains longer in mathematics than in physics. In physics, mathematics, as well as in astronomy and astrophysics, arXiv versions are cited more promptly and decay faster than WoS papers. The arXiv versions of papers—both published and unpublished—have lower citation rates than published papers, although there is almost no difference in the impact of the arXiv versions of both published and unpublished papers.

**Introduction**

Preprints—"temporary documents whose function is to bridge the time-gap created by publication delays" (Goldschmidt-Clermont, 1965, p. 8)—are a well-established mechanism for the exchange of scientific information (Mikhailov, Chernyi, & Giliarevskii, 1984). This is particularly true in astronomy and physics, disciplines that have long used preprints to communicate research results (Brooks, 2009; Brown, 2001; Kling, 2005; Wilson, 1970) and establish priority claims, thereby effectively reducing the role of the journal to "secondary distribution, archiving, and peer review" (Brooks, 2009, p. 92). Advocates of open access view subject repositories, such as arXiv, as heralding the eventual demise of the scholarly journal and have outlined ways in which peer review might function on these new platforms (Rodriguez, Bollen, & Van de Sompel, 2006), while others look forward to "the stranglehold journal publishers have over science libraries" being broken (Carriveau, 2008, p. 73). Hence the question: can or need these two forms of scholarly communication co-exist (Morris, 2003)?

First a word of caution: one should not be blinded by enthusiasm for the new. Preprints, after all, are far from novel. By way of illustration, preprints were cited as far back as 1922 in a Physical Review paper[1], and the Information Exchange groups, run by the National Institutes of Health, circulated already more than 1.5 million preprints in the mid-sixties (Confrey, 1996). Moreover, relatively few scholars, with physicists and mathematicians being notable exceptions, use preprints extensively (Swan & Brown, 2003). Lastly, what appears to work as a publishing model in one field may not translate to another (Kling, Spector, & Fortuna, 2004; Kling, Spector, & McKim, 2002).

Since its creation by Paul Ginsparg in 1991, arXiv has become central to the diffusion of research in a number of related fields, physics, mathematics, and computer science in particular (Gentil-Ceccot,

---

[1] Mckeehan, L.W. (1922). The crystal structure of silver-paladium and silver-gold alloys. Physical Review, 20(5), 424- 432.

Mele, & Brooks, 2009). Previous research has examined the use made of arXiv (Brown, 2001), ordering and citation rates (Dietrich, 2008a, 2008b; Haque & Ginsparg, 2009), the coexistence of e-prints and journals (Henneken et al., 2007), and the effect of arXiv on citation rates (Moed, 2007). However, data from all of arXiv and the Web of Science (WoS) database have yet to be combined for a comparative analysis. This paper combines the entire arXiv repository with the entire WoS database in order to better understand the ecology of scholarly communication. More specifically, we investigate (a) the proportion of papers across all disciplines that are on arXiv, (b) the elapsed time between arXiv submission and journal publication, and (c) the aging characteristics and scientific impact of arXiv e-prints and their alter egos (the versions published in the journal of record). This last analysis is performed on a subset of the dataset comprising papers published in astronomy and astrophysics.

**Background**

*Terminology*

John Ziman (1968) defined a preprint as:

> A clumsy, bulky, semi-legible document, being a duplicated version of a paper submitted for publication but not yet accepted and printed. It is a mechanized version of the decent and proper custom of writing to one's friends, colleagues and rivals about one's current work. (p. 110)

In other words, a preprint is something that has been submitted for publication, but has not yet been printed—something "intended for submission" (McKiernan, 2000, p. 127), even if it has not yet been submitted (Youngen, 1998). This raises the question of whether a "never-published manuscript" can be considered a preprint (Kling, 2005, p. 598). Kling (2005) argues that preprints should strictly refer to articles that have been accepted at a specific venue, while Brown (2001) stresses that a "preprint is the precursor to an article that *may eventually* be published in a peer-reviewed journal" [italics added] (p. 187). Increasingly, a preprint is submitted to a repository at the same time it is submitted to a journal, or after it has been accepted or even published, thus expanding the notion of a "pre" print to include refereed and post-print works (Brooks, 2009; Kling, 2005; Pinfield, 2009).

The transformation from print-based to electronic publishing has resulted in terminological conflation: e-print with pre-print (Brown, 2001)[2]. (The term e-print has been used inclusively to refer to both pre- and post-prints in the electronic environment (Brody, Harnad, & Carr, 2006)). Most broadly, e-prints have been defined as "scholarly and professional works electronically produced and shared by researchers with the intent of communicating research findings to colleagues" (U.S. Department of Energy, 2012). The idea of publication is central to both the conceptualization of a preprint server and the discussion of its relative merits and disadvantages, but is far from straightforward (Kling & McKim, 1999). As Eysenbach (2000, p.500) asks, "how many readers are needed to constitute 'publication'?" Kling (2005) argues that publishing is a "continuum rather than a binary (yes/no) proposition," and suggests that publication can "range from a one-day posting on a Web site to appearance in print in a large circulation, prestigious, peer-reviewed scientific journal" (p. 594). Kling (2005) also claims that a scholarly document is "published" when it has satisfied three criteria: publicity, trustworthiness, and accessibility (p. 594). Note that Kling suggests that posting to the Internet makes something public (and, therefore, published). However, Kling, Spector, and McKim (2002) qualify this by adding that something could be "weakly" published, in that it has limited publicity, trust, and accessibility.

---

[2] For a comprehensive history of paper precursors to e-print repositories, see Kling (2005)

*Advantages and disadvantages*

Enforcement of the so-called Ingelfinger rule—that is, rejecting an article due to prior publication of the same work—was something of a deterrent to the growth and development of preprint servers, as many communities lacked consensus on whether the public dissemination of a preprint constituted prior publication (Butler, 2001; Eysenbach, 2000). In early studies, most journal editors reported no formal policies on previously published work, although they distinguished between items that were previously published in journals and conferences and those appearing in less formal venues (i.e., the author's website, preprint archives, listservs), favoring publication of the latter (Harter & Park, 2000). However, there has been a marked relaxation of the Ingelfinger rule in recent years as use of the Internet has become ubiquitous and 'green' open access archiving has become increasingly accepted by commercial publishers (e.g., Elsevier).

Even so, there are outstanding issues and concerns. As Glaze (1999) noted:

> Electronic prepublication is not in the same category as the practice of sending out a few preprints to one's friends and colleagues. It reaches an audience that is larger, literally by orders of magnitude, compared to traditional preprint sharing. Indeed, electronic preprint publication is essentially a substitute for peer-reviewed journal publication in one respect, getting to the consumer. (p. 265)

The question of whether the entire intended audience is satisfied with an electronic preprint remains (Kling, 2005, p. 593). Another concern with preprint servers has to do with the dissemination of manuscripts that have not been peer-reviewed (see Kling [2005] for a discussion of controversies around communication of unrefereed manuscripts). This is particularly problematic in "medicinal, pharmaceutical, and biological chemistry where erroneous information can have life threatening implications" (Brown, 2003, p. 368). Skeptics argue that peer review "efficiently winnows out the wheat from the chaff" (Delamaothe, Smith, Keller, Sack, & Wischer, 1999, p. 1516) and prevents the academic community from being drowned in "junk science" (Carriveau, 2008, p. 77). Eysenbach (2000) counters by arguing that with preprints the focus "is not on validity (the reader is aware that he is dealing with draft data) but on openness and speed" (p. 501), while Ginsparg (1994) believes that peer review is superfluous, given that "the fear of ruining one's professional reputation will prevent someone from submitting poorly developed manuscripts to an e-print archive" (c.f. Carriveau, 2001, p. 78). Advocates of preprint servers also note the accelerating pace of scientific discovery and the progressive democratization of scholarly communication (Eysenbach, 2000; Gentil-Beccot, Mele, & Brooks, 2009).

*arXiv and related platforms*

Paul Ginsparg launched xxx.lanl.gov, the first Internet-based e-print server, in 1991 to facilitate preprint exchange in the field of theoretical high-energy physics (Brown, 2010; Carriveau, 2008; Davis & Fromert, 2007; Ginsparg, 2008). The name was changed to arXiv.org in 1998, after the service grew in popularity and expanded to cover other fields (Ginsparg, 2008). The aim was to create an electronic bulletin board "to serve a few hundred friends and colleagues" (Ginsparg, 2011, p. 145) and to "automat[e] a paper-based process already in existence" (Pinfield, 2001, para. 1). Today's much-enlarged arXiv is strongest in physics, mathematics, and computer science (Brody, Harnad, & Carr, 2006), fields in which there is a tradition of preprint use.

The number of articles in arXiv has been growing linearly since 1991, and arXiv is now the "largest self-archived centralized e-print archive" (Brody, Harnad, & Carr, 2006, p. 102). Originally hosted at the Los Alamos National Laboratory (hence the initial domain name), it was later moved to Cornell University, where it is currently under the aegis of the university library (Hey & Hey, 2006). In parallel, Ginsparg began a movement to develop a set of technical standards for the establishment of a global preprint archive via the Universal Preprint Service Initiative—later known as the Open Archives Initiative (OAI) (Brown, 2001; Manuel, 2001). The federated nature of OAI repositories has led to proposals for a "repository-centric peer-review model" based on the OAI platform and using a social-network algorithm to suggest potential reviewers and weigh evaluations (Rodriguez, Bollen, & Van de Sompel, 2006).

In 1997, arXiv began collaborating with the Astrophysics Data System (ADS) and the ADS created an index for astrophysics e-prints, making them available through the ADS abstracts service. In 2002, abstracts of all arXiv categories were included (Henneken et al., 2007). arXiv also has a relationship with SPIRES, the first electronic catalogue of grey literature, focused on high-energy physics preprints (Gentil-Beccot, Mele, & Brooks, 2009). SPIRES counts citations to and from preprints and directs physicists to arXiv (82% of clicks from SPIRES go to arXiv) (Gentil-Beccot, Mele, & Brooks, 2009). SPIRES is currently being replaced with INSPIRE, which was created to "provide an even more flexible and extensible system to allow publishers, repositories, and researchers themselves to contribute and share information" (Brooks, 2009, p. 91). A survey of high-energy physicists found that nearly 90% rely on SPIRES and arXiv as their point of entry to the literature. This system is so embedded in the working practice of physicists that Kling, McKim, and King (2003) considered SPIRES, arXiv, and associated human actors as the embodiment of a functioning socio-technical interaction network.

*Other preprint repositories*

Numerous other preprint repositories have been created, although few have been as successful as arXiv. In the mid-1990s, the American Mathematical Society (AMS) sponsored an e-print repository for mathematics. By 1999, the AMS suspended operations and endorsed arXiv for mathematics e-prints (Kling, 2005). Patrick Brown of Stanford University led an initiative to create a preprint archive for biologists (Butler, 2001), and biology was added to arXiv as a subject area in 2003 (Butler, 2003). The ACM similarly partnered with arXiv rather than developing its own server for computer science preprints (Kling, 2005). The Chemistry Preprint Server, launched in 2000 (Brown, 2003), was terminated four years later (Brown, 2010). For the social sciences and humanities, the Social Sciences Research Network contained, as of June 10th 2013, close to half a million abstracts as well as the full-text of about 400,000 papers[3]. A list of other, specialized subject e-print repositories can be found in Garner, Horwood, and Sullivan (2001).

Preprint repositories have also been created at the national level: the DOE launched the e-print network (Traylor, 2001), a gateway to more than 35,500 websites and databases, containing 5.5 million e-prints in basic and applied sciences (U.S. Department of Energy, 2012). The Ministry of Education in the People's Republic of China created the Chinese Science Paper Online (CSPO) in 2003, which includes a government requirement that authors receiving funding from the government submit at least two working papers (Hu, Zhang, & Chen, 2010). Despite these mandates, preprint servers in China still have relatively limited impact (Yaokun & Nanqiang, 2008).

*Empirical investigations of arXiv*

Over the years, several studies have focused on authors' practices with respect to arXiv: Fowler's (2011) survey of mathematicians found that 81% had posted to arXiv and that it was a regular sharing mechanism for 30%; Manuel (2001) found that authors were primarily academic (rather than corporate); and Moed (2007) showed that posters tended to be high-impact authors (measured by the citation impact of those of their papers *not* deposited in arXiv). However, most research has focused on the preprints—specifically on the relationship between electronic preprints and their subsequent publication and impact. For example, approximately half of all preprints in arXiv subsequently appeared in peer-reviewed publications (Manuel, 2001; Mine, 2009), and Moed (2007) found this percentage to be about 75% in condensed matter. Studies have also looked at the inverse, viz., the proportion of journal literature in a given field that is also in arXiv. The rate was almost 100% in high-energy physics (Gentil-Beccot, Mele, & Brooks, 2009) but 18.5% in mathematics (Davis & Fromerth, 2007). The number of articles appearing in both arXiv and the published literature is increasing (Davis & Fromerth, 2007; Gentil-Beccot, Mele, & Brooks, 2009). Peer-reviewed articles that were also preprints receive significantly more citations than articles not deposited (Davis & Fromerth, 2007; Gentil-Beccot, Mele, & Brooks, 2009). The reasons suggested include an early view effect, a quality differential, and an open access advantage (Davis & Fromerth, 2007; Kurtz et al., 2005).

---

[3] http://www.ssrn.com/

Some studies confirm the early view effect: "colleagues in the field start the process of reading a paper, processing its information, and citing it in their own articles earlier if a paper is deposited in arXiv" (Moed, 2007, p. 2053). However, other studies have found no such effect (Davis & Fromerth, 2007). Evidence has also been found to support a "quality bias"; that is, better papers and high impact authors appear in arXiv more than the reverse (Davis & Fromerth, 2007; Moed, 2007). Little or no support has been found for the open access advantage, however (Davis & Fromerth, 2007; Kurtz et al., 2005; Moed, 2007). As Kurtz et al. (2005) concluded, "there is no significant population of astronomers who are both authors of major journal articles and who do not have 'sufficient' access to the core research literature" (p. 1400-1401). Haque and Ginsparg (2009, 2010) found that posts on arXiv at the beginning and end of the day receive higher levels of citation and readership than those in the middle. Other studies have examined the proportion of citations to the e-print version of the paper, with mixed findings (Manuel, 2001; Youngen, 1998).

Readership has also been investigated. Using two years of cumulative download and citation data from arXiv, Brody, Harnad and Carr (2006) found that download counts at six months provided reliable predictions of citation impact at two years. They concluded that "the rapid dissemination model of arXiv has accelerated the read-cite-read cycle substantially" (p. 1062). The relationship between the publisher's version and the preprint remains unclear: Davis and Fromerth (2007) found that arXiv-deposited articles received 23% fewer downloads from publishers' websites. However, in a study of four astronomy journals, Henneken et al. (2007) found that reads of the arXiv e-print through ADS dropped to zero (or near zero) immediately following the publication of the peer-reviewed article. They also note that the half-life of e-prints is shorter than that of the corresponding journal articles, concluding that, "e-prints have not undermined journal use in the astrophysics community and thus do not pose a threat to the journal readership" (Henneken et al., 2007, p. 19).

All in all, this literature shows that e-prints are having an effect on how scientists communicate the findings of their research. However, the precise nature of the effect(s) remains fuzzy. Our study, by comparing comprehensive arXiv and WoS datasets, should lead to greater clarification and insights.

**Methods**

Here we use two data sources: the arXiv database and WoS. All arXiv database metadata from 1990 to March 22, 2012 were downloaded (n=744,583 e-prints). All standard citation indexes were used for WoS (Science Citation Index Expanded, Social Sciences Citation Index and Arts and Humanities Citation Index) for the 1990-2011 period. Data are presented for 1995—2011 (or 2010 in some cases), although citations and matching papers were compiled until the end of 2012. Two types of links between the data sources were created: (a) between the arXiv e-print and its published version indexed in WoS, and (b) between the arXiv e-print and the citations it received in WoS. Several steps were needed to match the arXiv e-print to its published counterpart (a). First, three sets of links were established: 1) direct correspondence between the arXiv and WoS titles; 2) fuzzy matching between the arXiv and WoS titles AND fuzzy matching between the journal mentioned in the arXiv bibliographical notice and the WoS journal; and 3) fuzzy matching between the arXiv and WoS titles AND fuzzy matching between the arXiv first author and the WoS paper first author. These links were, in a second step, automatically validated through the similarity between the respective abstracts of WoS papers and arXiv e-prints. In order to reduce the computing time of such validation steps, the similarity was computed using the first 20 to 50 characters from the abstracts. More specifically, for papers that had apparently matching titles and authors / journals, the correspondences were validated using the first 20 characters of the abstracts or the first 30 following the first period (.) for abstracts that are divided into sections. For the abstracts not divided into sections, the validation was performed using the first 50 characters of the abstracts. After this round of matching – which yielded about 440,000 pairs of arXiv and WoS documents – we expanded the matching criteria to 1) the same first author, 2) a publication delay between arXiv submission and WoS from -1 to +5, 3) journals that have published papers submitted to arXiv and 4) titles having at least 90% similarity (irrespective of both documents' numbers or authors) or 60% similarity when the number of authors was the same. Finally, a last round of matching was performed using the DOIs of papers and DOIs found on arXiv, which added 13,129 documents to the match.

In total, 474,011 out of the 744,583 arXiv e-prints (63.7%) were matched with a WoS-indexed journal article, note, or review. Hence, an arXiv preprint that links to an editorial, letter to the editor or to other non-peer-reviewed material is not included. A paper that has the *same* scientific content as an e-print, but that has had significant change in its authorship, title or abstract, was not considered to be the same document. In other words, our methods allow for the matching between two documents having the same "bibliographic" properties, but not for the matching of documents with the same content. Thus, they underestimate the proportion of arXiv's *scientific content* that is found, at some point, in WoS-indexed journals.

For the second matching (b) we utilized the specific structure of the references to the arXiv e-prints in WoS. For example, a reference to an e-print from the condensed matter section of arXiv will have the string "CONDMAT" followed by the series of seven or eight digits that correspond to its document ID in the online e-print database. Given that a paper belonging to more than one arXiv category can be cited using both categories as prefixes, the matching process used the seven or eight digits as well as its prefix. For astronomy and astrophysics, we separated documents into four distinct categories: 1) arXiv e-prints never published in a WoS-indexed journal, 2) arXiv e-prints published in a WoS-indexed journal, 3) WoS-indexed journal articles also published and archived as an arXiv e-print, and 4) WoS-indexed journal articles that were never published as arXiv e-prints. Finally, the field classification used is that of the U.S. National Science Foundation,[4] developed by *The Patent Board*.

**Results and discussion**

*Proportion of ArXiv e-prints published in WoS journals*

Figure 1 presents the proportion of arXiv e-prints published as papers and indexed in the WoS, by arXiv subject classification. It shows that, globally, about 64% of arXiv can be found in the WoS. This percentage is slightly higher than those obtained by Manuel (2001) and Mine (2009). This percentage varies across subfields: while about 80% of e-prints in condensed-matter physics (physics:cond-mat), and 70% of those in theoretical nuclear physics (physics:nucl-th), and theoretical high-energy physics (physics:hep-th) are published in WoS journals, this percentage is around 45% in mathematics (math), quantitative finance (q-fin), and statistics (stat), and is less than 20% in computer science (cs). This could be explained by the (other) preferred modes of diffusing research in those disciplines: while physicists are users of the scientific paper (Larivière, Archambault, Gingras & Vignola-Gagné, 2006), computer scientists rely heavily on conference proceedings (Lisée, Larivière & Archambault, 2008), hence their lower proportion of e-prints being indexed in the WoS.

---
[4] http://www.nsf.gov/statistics/seind06/c5/c5s3.htm#sb1

A striking aspect of Figure 1, found in the Inset, is the stability of the percentage of arXiv e-prints that are published in WoS journals (with the exception of the most recent years, which is likely a consequence of the delay between arXiv submission and publication as well as the increase in submissions from disciplines such as mathematics and statistics where a lower proportion of arXiv papers make it to WoS journals). Indeed, between 1995 and 2006, more than 73% of e-prints make it to a WoS journal and, despite the increase in WoS coverage over the recent years, this figure remains remarkably stable, which suggests that users of arXiv have, from the beginning, used both diffusion platforms *in the same proportions* as they are used today. Similarly, other modes of diffusion—outside WoS journals or in conference proceedings, for instance—or the consideration of the arXiv e-prints as the "final" format of diffusion of the paper have also been stable, accounting for slightly more than 25% of all e-prints.

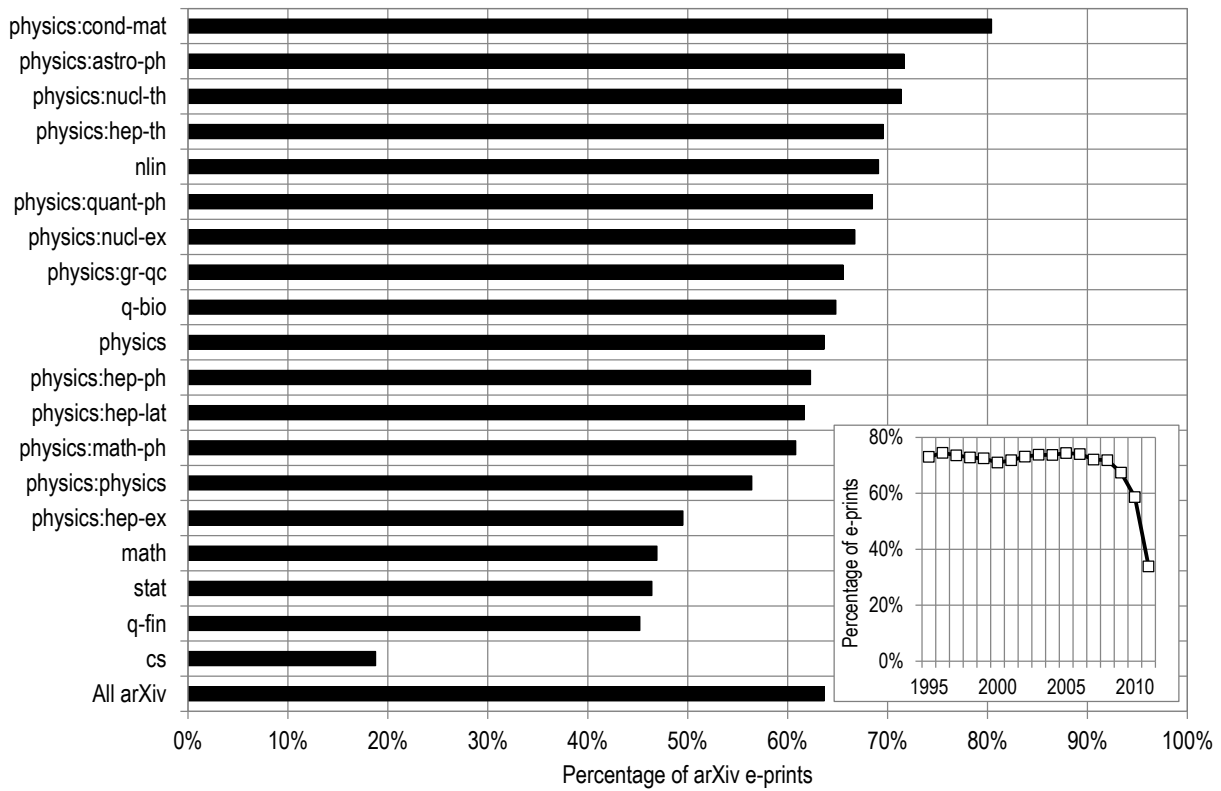

**Figure 1. Proportion of arXiv e-prints published in WoS-indexed journals, by arXiv specialty (1995-2011). Inset: Evolution of the proportion arXiv e-prints published in WoS-indexed journals (1995-2011).**

*Proportion of WoS papers on arXiv*

As mentioned previously, about 64% of all arXiv e-prints are published in a WoS-indexed journal. However, when WoS papers are taken as the denominator, only 3.6% of 2010 WoS papers (all disciplines combined) were submitted to arXiv (Inset of Figure 2). Figure 2 also shows that three disciplines account for the vast majority (93%) of arXiv submissions in 2010-2011: mathematics (with 21% of all WoS papers on arXiv), physics (20% of all WoS papers on arXiv) and earth and space (12% of all WoS papers on arXiv). Within these disciplines, a few specialties are using it more intensively.

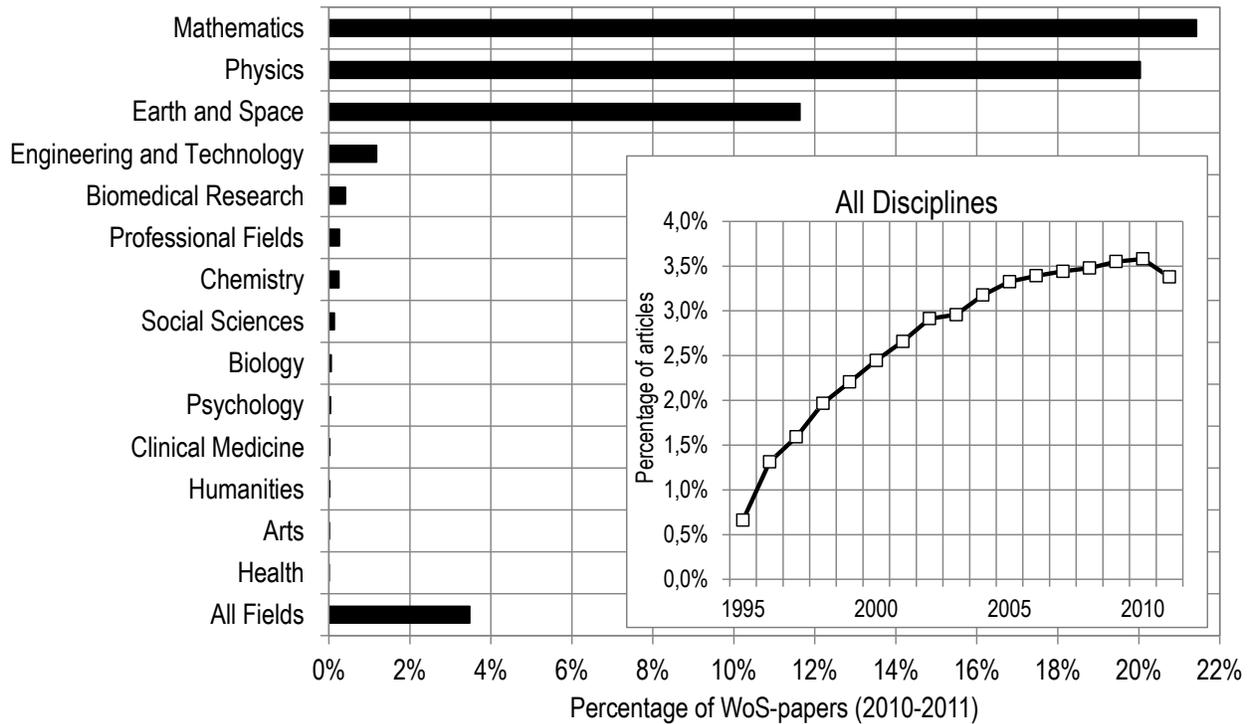

**Figure 2. Proportion of WoS papers on arXiv, by specialty (2010-2011). Inset: Proportion of WoS papers on arXiv, by specialty, 1995-2011.**

As shown in Figure 3, about two-thirds of WoS papers published in astronomy, astrophysics, and nuclear and particle physics are found on arXiv. The Inset of Figure 1 shows that this percentage has increased since 1995. While researchers in nuclear and particle physics were quick to adopt arXiv—this percentage was already at 63% in 2000—those in astronomy only gradually made higher use of it. Since the mid-2000s, both specialties have used arXiv to the same extent. In nuclear and particle physics, the percentage we obtain is lower than that of Gentil-Beccot, Mele, and Brooks (2009) for high-energy physics, which is due to the fact that their definition of the field only included 5 high-impact journals, while ours covered 48 journals including nuclear physics journals. In mathematics, our percentages are higher than those of Davis and Fromerth (2007), which is likely a consequence of the increasing number of papers appearing in both arXiv and in the WoS.

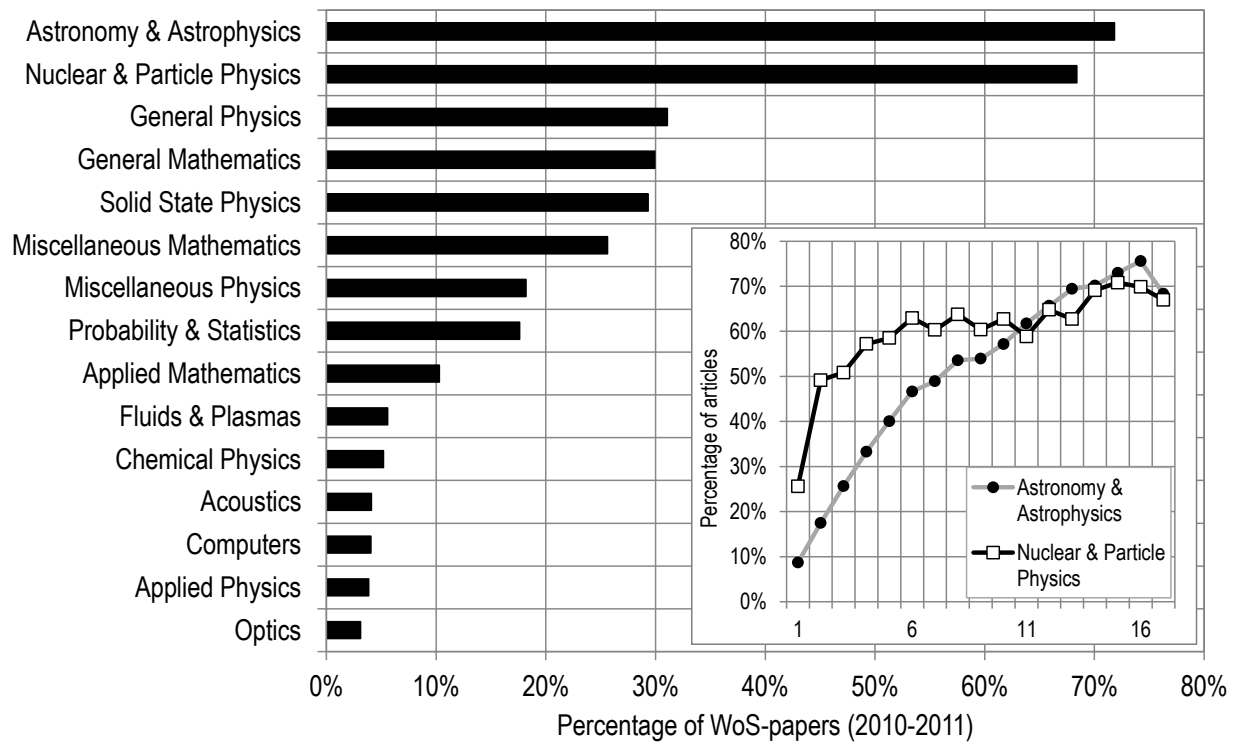

**Figure 3. Proportion of WoS papers on arXiv, by specialty (2010-2011). Inset: Proportion of WoS papers on arXiv, by specialty, 1995-2011.**

Figure 4 shows that the elapsed time between the submission of the manuscript to arXiv and publication in a peer-reviewed journal has decreased.[5] Whereas papers were once published a year after appearing on arXiv, publication in a journal is now likely to occur in the same year as the paper's appearance on arXiv. Although we do not have a clear explanation for this, it might be due to new technologies that make the publication process quicker, the emergence of new fast-publishing journals like PLOo ONE and Nature Scientific Reports, a higher proportion of researchers waiting for a paper to be published or accepted for publication before submitting it to arXiv, or the introduction of arXiv, which may have motivated publishers to try to reduce publication delays.

---

[5] 11,946 e-prints out of 440,371 that matched to a WoS paper (2.7%) have been submitted on arXiv after journal publication; those have been removed from this part of the analysis.

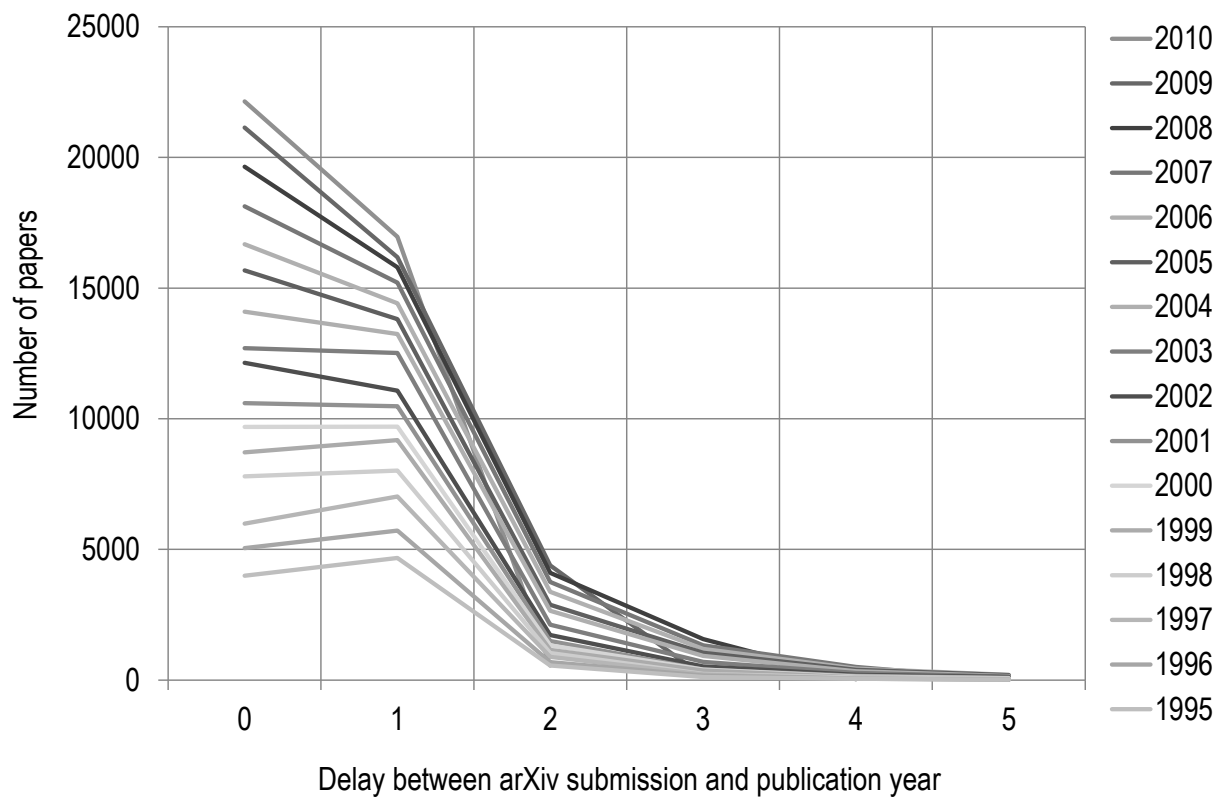

**Figure 4. Distribution of the elapsed time between arXiv submission and publication year, by year of submission to arXiv, 1995-2010.**

Elapsed time between arXiv submission and journal publication varies dramatically across specialties of science. Figure 5 presents this interval—compiled as an average—for the top 18 specialties with the highest number of WoS papers found on arXiv. It globally shows that physics specialties have very short delays—less than half a year on average—while those of mathematics have longer delays (>1 year). Among the specialties with the shortest time between arXiv submission and journal publication is the category of astronomy and astrophysics, one of the two specialties with the most intensive use of preprints. The appearance of "general biomedical research" is due to the fact that "general" journals that publish physics or mathematics papers, such as *Science* and *Nature,* are assigned to this category.

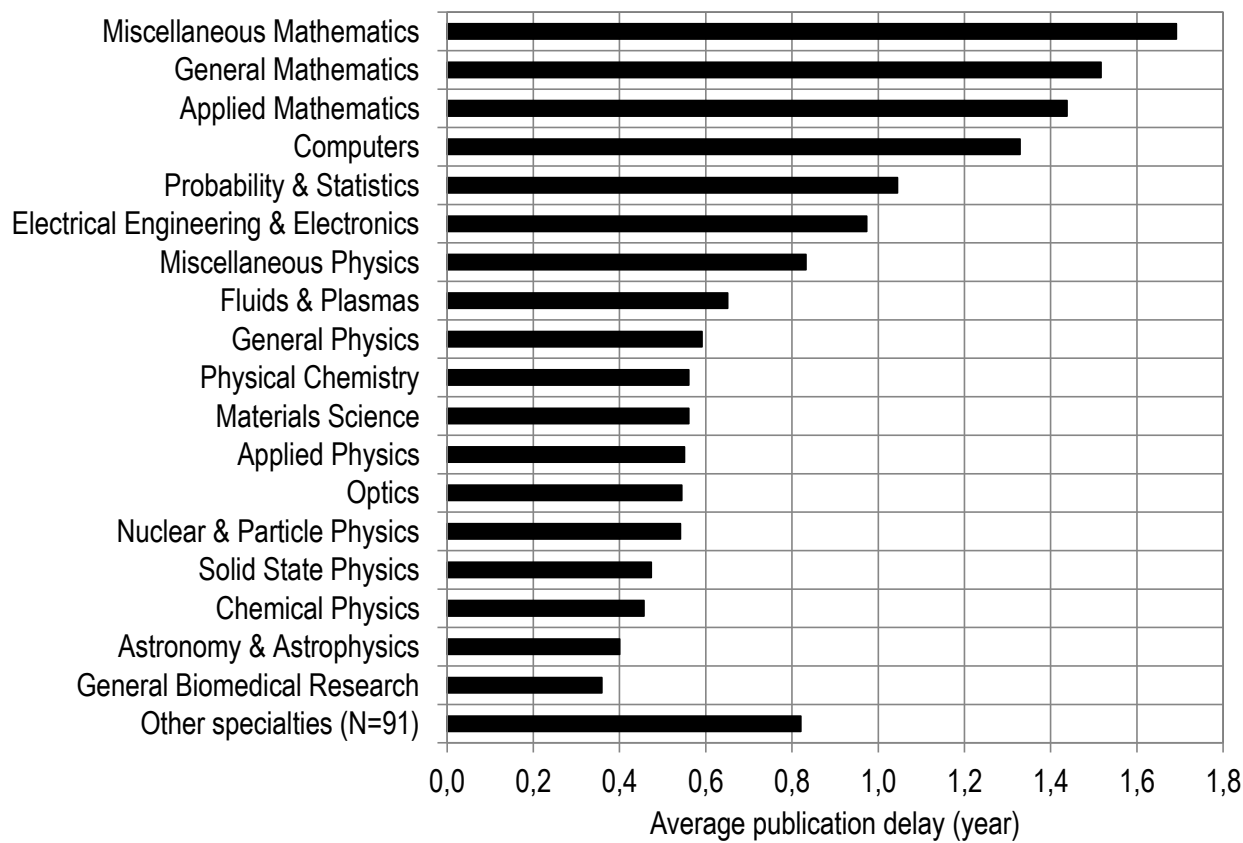

**Figure 5. Average elapsed time between arXiv submission and journal publication, by specialty (top 19 specialties with the highest number of WoS papers found), 1995-2011.**

*Proportion of WoS papers' cited references made to arXiv e-prints*

Figure 6 presents the proportion of WoS papers' cited references that are made to arXiv e-prints, by discipline (panel A) and specialty (panel B). At the level of disciplines, physics has the highest proportion (1.4% of references in 2011), followed by mathematics (1% in 2011) and earth and space science (0.2% in 2011). At the level of specialties, journals in nuclear and particle physics are in a league of their own, with 6.6% of their references made to arXiv e-prints. In the other specialties with the highest proportion of references made to arXiv (mainly physics and mathematics), this percentage is below 1.5%. Worth noting is the percentage of references made to arXiv in one specialty from the social sciences, "science studies," which is likely due to a greater use of arXiv by scholars in the bibliometrics community.

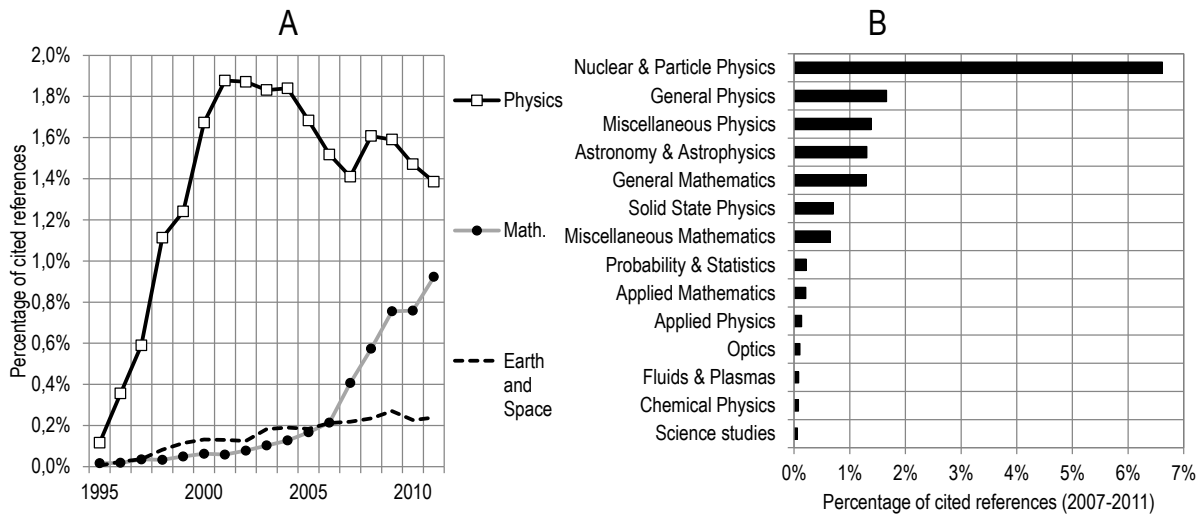

**Figure 6. A) Proportion of WoS-papers' cited references made to arXiv preprints for the top three disciplines, 1995-2010 B) Proportion of WoS-papers on arXiv, for top specialties (2007-2011).**

All in all, these numbers show that e-prints *are* cited, and the results are surely underestimations. As discussed by Brown (2001, pp. 192-193), editors were not enthusiastic about citing e-prints in published manuscripts, and expressed concern regarding preservation and peer-review. At the time of her analysis, none of the journals of the American Institute of Physics allowed subsequent citation to e-prints (it appears that they have since changed this policy). Similarly, the fact that editors would "encourage" authors to "update the reference if the work [was] published" during the refereeing process and that authors were "cautioned that references to preprint and other sources that are not readily available to readers should be avoided" suggests that our results underestimate the impact of arXiv e-prints that have a *published* alter-ego. In addition, we likely underestimated the impact of arXiv e-prints due to the fact that references made in these documents were not available to us; it would be reasonable to presume that there would be a higher proportion of references made to other arXiv e-prints in arXiv e-prints. Larivière, Archambault, and Gingras (2008) suggested that the introduction of arXiv might have been responsible for the decrease in the average age of cited literature in astrophysics and related areas. Figure 7 tends to confirm this hypothesis: in all disciplines, cited arXiv e-prints are significantly younger than all cited material taken together. At the level of all disciplines, cited e-prints have an average age of 2.2 years, while the average is 7.1 years for all cited documents combined. On the other hand, as shown by Milojević (2012), the effect of e-prints of the age of cited literature has been transitory, until the use of e-prints became widespread.

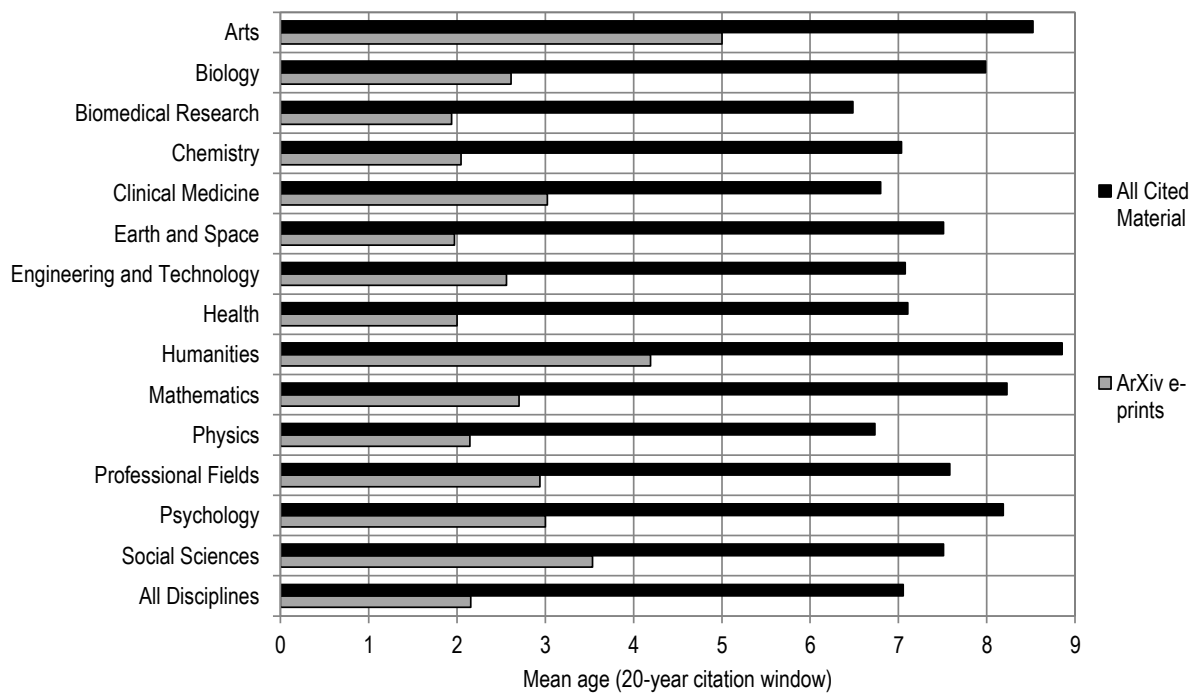

**Figure 7. Mean age of all cited documents and of cited e-prints, by discipline, 1995-2011.**

*Aging characteristics and scientific impact*

This section analyzes the aging characteristics and scientific impact of arXiv e-prints and their WoS-published alter egos for physics and mathematics, as well as for the specialty of astronomy and astrophysics. Figures 8 and 9 present, for physics and mathematics (taken as whole) respectively, the trends in the numbers of papers that have appeared on arXiv only, on arXiv and WoS (arXiv version), in WoS only, and on arXiv and WoS (WoS version) (panel A), and the mean number of citations these documents have received using a one-year citation window, including publication year (panel B). When one takes the sum of all distinct documents (WoS + arXiv) as the denominator, it shows that, in both physics and mathematics, there is a decline of papers found in WoS. More interestingly, perhaps, are the citation rates of the four groups of papers. In both disciplines, papers that have the highest (short-term) impact are the WoS versions of papers that are also arXiv e-prints, a finding consistent with the well-documented association between arXiv submission and citation (Davis & Fromerth, 2007; Gentil-Beccot, Mele, & Brooks, 2009). Although the difference between WoS papers that are also arXiv e-prints and WoS papers that are not is relatively stable in physics, the gap between the two is decreasing in mathematics. arXiv versions—both published and unpublished—obtain lower citation rates. Surprisingly, however, there is almost no difference in the impact of the arXiv versions of published and unpublished papers. One could have expected that these unpublished papers, being non-refereed, would have a lower impact than comparable arXiv submissions published in a journal. However, it is possible that researchers prefer to cite the published version of an e-print, which is likely to reduce published e-print impact and hence make the two measures comparable. On the whole, these results are consistent with those of Brooks (2009), who showed that unpublished arXiv submissions had five times less impact than those published in a journal, when one includes the citations received by the published version of the e-print.

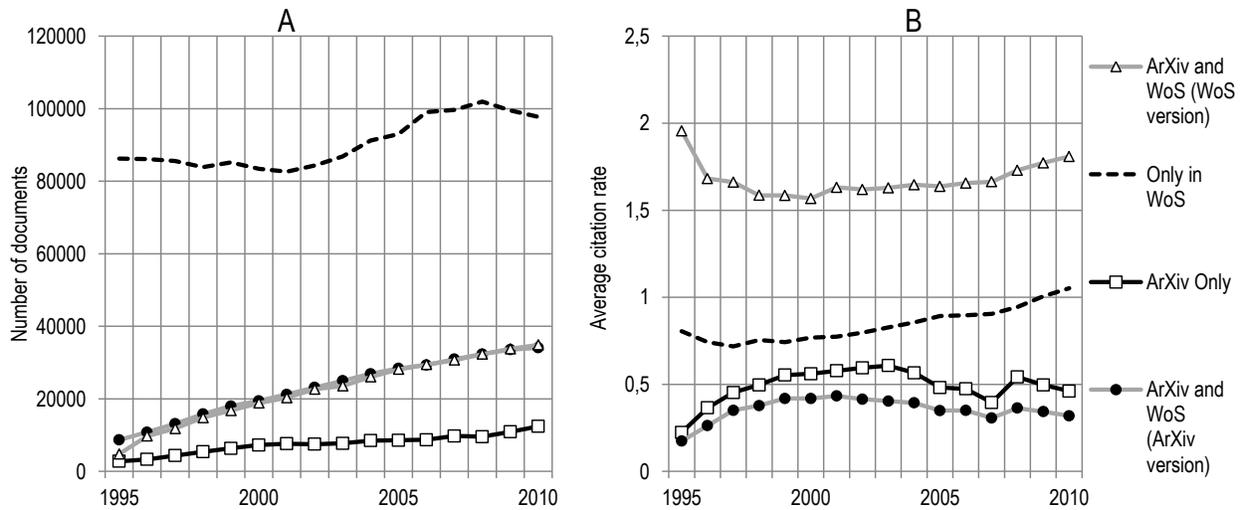

**Figure 8.** For physics A) Number of documents published and B) mean number of citations received (publication year plus one year), for documents published on arXiv only, on arXiv and WoS (arXiv version), only in WoS and on arXiv and WoS (WoS version), 1995-2010

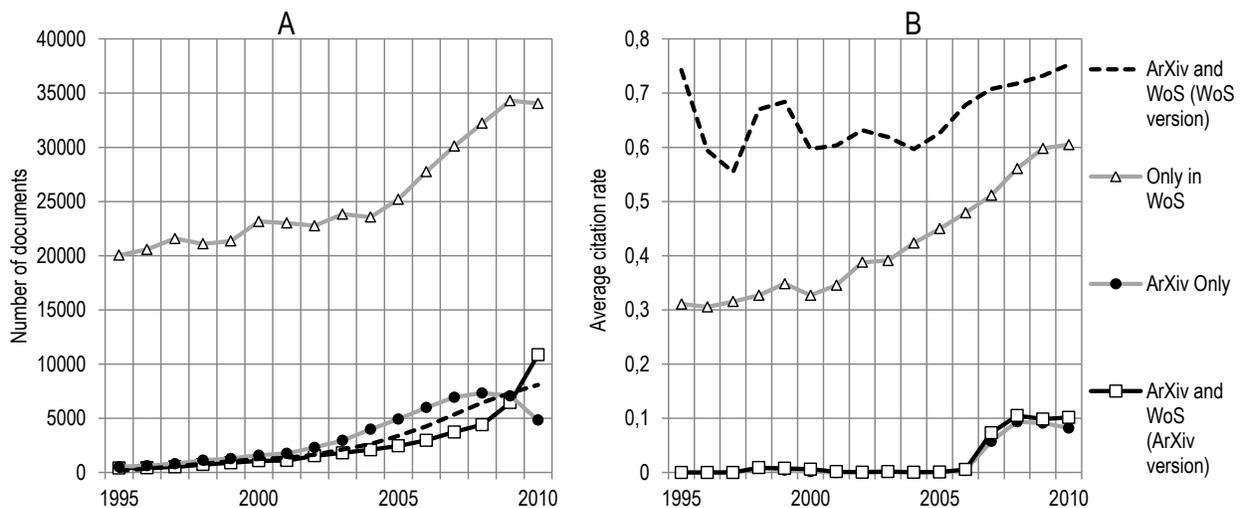

**Figure 9.** For mathematics A) Number of documents published and B) mean number of citations received (publication year plus one year), for documents published on arXiv only,

on arXiv and WoS (arXiv version), only in WoS and on arXiv and WoS (WoS version), 1995-2010

For the subset of astronomy and astrophysics papers and e-prints (Figure 10), we see a considerable increase in the number of documents published both in arXiv and in journals, a small increase in the number of papers published only in arXiv (or outside WoS-journals), and a decline in papers published only in journals. In terms of proportion of all distinct astronomy and astrophysics documents in 2010—obtained by the combination of arXiv and WoS—16% are found on arXiv only, 59% can be found both on arXiv and in WoS, and 25% are found in WoS only. The citation rates of the four groups of documents are quite different and vary over time. WoS versions of arXiv e-prints obtain the highest citation rates. However, this mean impact has initially decreased to remain quite constant since 1999—even when we add to the WoS version the citations received by the arXiv version—and is approaching that of other WoS papers not submitted to arXiv, whose mean impact is increasing. As with mathematics and physics, the impact of the arXiv versions of both published in WoS and unpublished (or published outside WoS) papers in astronomy and astrophysics is almost identical (although here it is slightly above in the case of arXiv only papers).

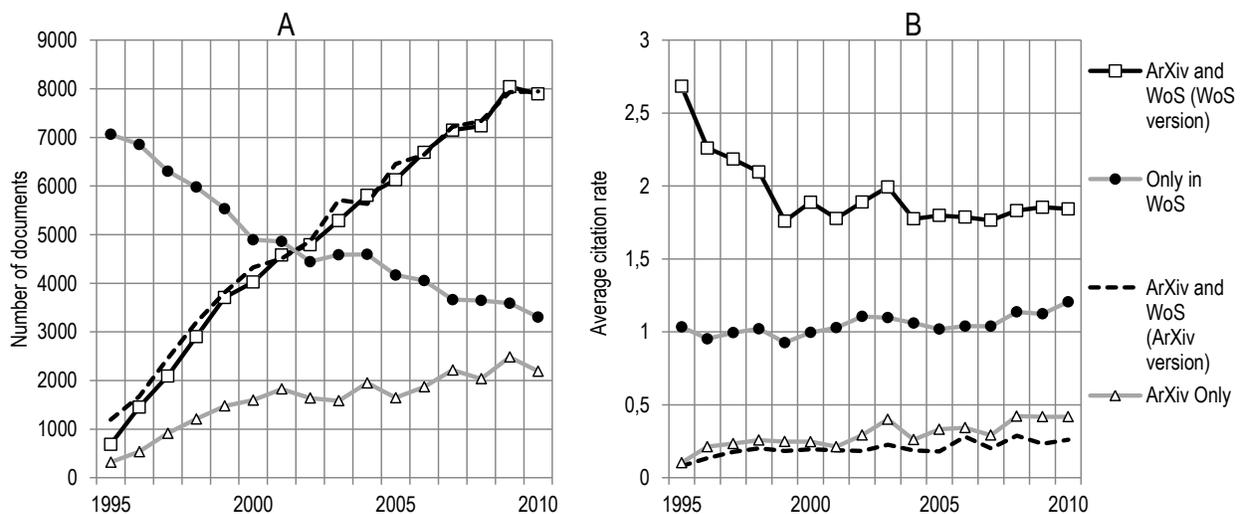

**Figure 10. For astronomy and astrophysics A) Number of documents published and B) mean number of citations received (publication year plus one year), for documents published on arXiv only, on arXiv and WoS (arXiv version), only in WoS and on arXiv and WoS (WoS version), 1995-2010**

In terms of aging characteristics, Figure 11 presents the age distribution of citations received by the four groups of documents. It shows that e-prints and published papers follow different patterns. Citations to e-prints peak the year following submission, while citations to papers published in journals are similar during the two years following the publication year. Given the transfer of citations from pre-publication e-prints to their published version (Brown, 2001; Henneken et al., 2007), citations to their e-print versions decay faster than those received by unpublished e-prints. E-prints found solely on arXiv have a slower decay, though not as slow as WoS papers. These faster citations for arXiv e-prints are consistent with the findings of Brody, Harnad, and Carr (2006) as well as those of Henneken et al. (2007).

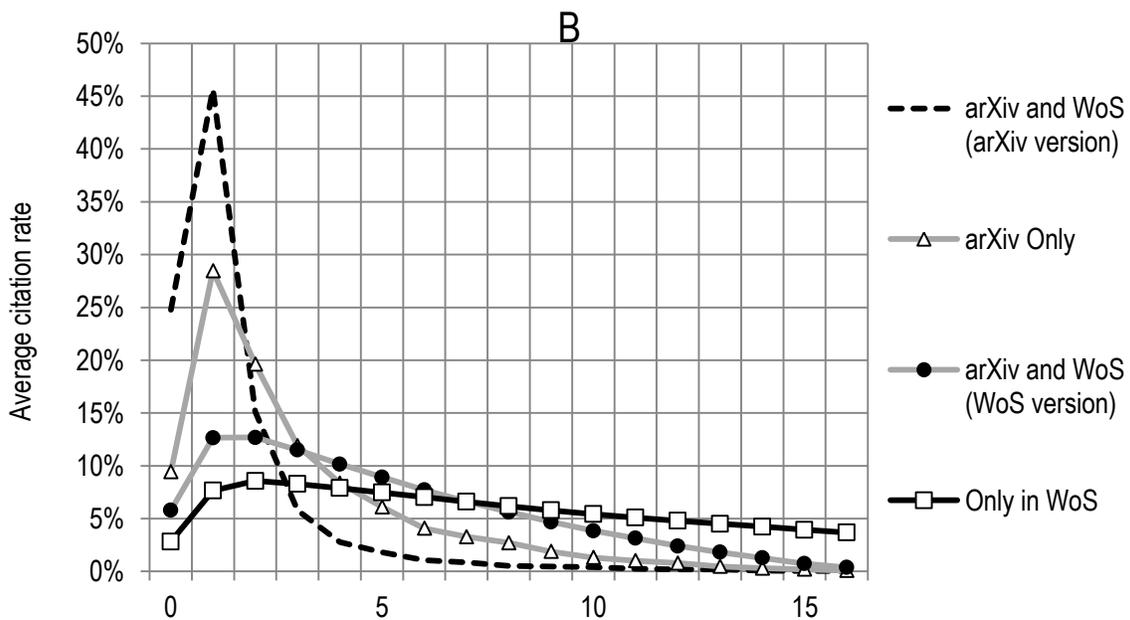

**Figure 11. Percentage of citations received, for documents published on arXiv only, on arXiv and WoS (arXiv version), only in WoS and on arXiv and WoS (WoS version), 1995-2010**

**Conclusion**

This paper shows that arXiv has changed the scholarly communication patterns of physicists and mathematicians. In some specialties, such as astronomy and astrophysics and nuclear and particle physics, the vast majority of papers published in WoS-indexed journals are found on arXiv. The role of arXiv in these communities has moved from the space for sharing pre-prints by the minority to the place for archiving the majority of produced research. However, in these disciplines there is still a significant number of papers that are not on arXiv. Previous research on the topic, focusing on high-impact journals exclusively, has found a greater proportion of WoS-papers in those specialties to be on arXiv (Gentil-Beccot, Mele, & Brooks, 2009). Our results show that, when the whole discipline is considered—high-impact and low-impact journals alike—the proportion of published papers that are self-archived on arXiv is noticeably lower. Similarly, not all specialties are using it to the same extent: in most physics and mathematics specialties, less than a third of WoS papers are found on arXiv. Along these lines, arXiv is increasingly used outside these two fields, but is still quite marginal: 93% of all WoS-published arXiv e-prints are published either in mathematics, physics, or earth and space sciences. Our results also show that the average elapsed time between submission to arXiv and publication in a WoS-indexed peer-reviewed journal has decreased over time. This may be due to new technologies that make the publication process quicker, to the emergence of new fast-publishing journals, to a higher proportion of researchers waiting for the paper to be published or accepted for publication before submitting to arXiv, or to a reduction in publication delays. These time lags are also quite different across fields of science, with physics specialties having shorter delays than mathematics specialties.

The subset of astronomy and astrophysics papers shows that arXiv versions of papers are cited more promptly and decay faster than WoS papers. WoS versions of arXsiv e-prints obtain the highest citation rates, but the difference with other WoS papers not submitted to arXiv is decreasing. Unsurprisingly, arXiv versions of papers—both published and unpublished (or published in non-WoS journals)—obtain lower citation rates, although there is almost no difference in the impact of the arXiv versions of both published and unpublished papers. As Brody, Harnad, and Carr (2006) point out, the fact that preprints are cited before publication (and, hence, peer review), as well as the fact that unpublished e-prints are cited, raises the question of the function of peer-review in those fields. It seems that citing authors either evaluate papers themselves, often being reviewers, or trust the results presented—which might be a consequence of the few massive collaborations and large-scale scientific infrastructures found in these disciplines.

So far, all the evidence suggests that the growth of e-pre-prints in mathematics and physics does not herald the demise of peer reviewed journal articles. Authors continue to communicate the results of their research using a mix of options and outlets, as our analysis, based on arXiv and WoS data, shows compellingly—albeit with differential rates of adoption across disciplines, even within the different specialties of physics. The ways in which scientists use arXiv is, thus, highly nuanced. What can be said with some assurance, however, is that arXiv performs an increasingly important role as both an alerting service and an archive of first resort, though it is equally clear that publication in a journal of record remains a demonstrably important goal for most scientists.

**Acknowledgments**

This research was funded by SSHRC (Canada), the National Science Foundation (U.S.; grant #1208804), and AHRC/ESRC/JISC (U.K.) through the *Digging into Data* funding program. VL also acknowledges funding from the Canada Research Chair program. The authors also wish to thank Michael Kurtz for detailed comments and suggestions.**References**

Brody, T., Harnad, S., & Carr, L. (2006). Earlier web usage statistics as predictors of later citation impact. *Journal of the American Society for Information Science & Technology, 57*(8), 1060-1072.

Brooks, T.C. (2009). Organizing a research community with SPIRES: Where repositories, scientists and publishers meet. *Information Services & Use, 29*, 91-96.

Brown, C. (2001). The e-volution of preprints in the scholarly communication of physicists and astronomers. *Journal of the American Society for Information Science & Technology, 52*(3), 187-200.

Brown, C. (2003). The role of electronic preprints in chemical communication: Analysis of citation, usage, and acceptance in the journal literature. *Journal of the American Society for Information Science & Technology, 54*(5), 362-371.

Brown, C. (2010). Communication in the sciences. *Annual Review of Information Science & Technology*, 287-316.

Butler, D. (2001). US biologists propose launch of electronic preprint archive.

Butler, D. (2003). Biologists join physics preprint club. *Nature*, 425.

Carriveau, K.L. (2008). A brief history of e-prints and the opportunities they open for science librarians. *Science & Technology Libraries, 20*(2-3), 73-82.

Confrey, E.A. (1996). The Information Exchange Groups experiment. *Publishing Research Quarterly, 12*(3), 37-39.

Davis, P.M., & Fromerth, M.J. (2007). Does the arXiv lead to higher citations and reduced publisher downloads for mathematics articles? *Scientometrics, 71*(2), 203-215.

Delamothe, T., Smith, R., Keller, M.A., Sack, J., & Witscher, B. (1999). Netprints: The next phase in the evolution of biomedical publishing: Will allow researchers to share their findings in full, for free, and fast. *BMJ: British Medical Journal, 319*(7224), 151-1516.

Dietrich, J.P. (2008a). The importance of being first: Position dependent citation rates on arxiv:astro-ph. *Publications of the Astronomical Society of the Pacific*, 120, 224–228.

Dietrich, J.P. (2008b). Disentangling visibility and self-promotion bias in the arxiv:astro-ph positional citation effect. *Publications of the Astronomical Society of the Pacific*, 120, 801–804.

Eysenbach, G. (2000). The impact of preprint servers and electronic publishing on biomedical research. *Current Opinion in Immunology, 12*, 499–503.

Fowler, K.K. (2011). Mathematicians' views on current publishing issues: A survey of researchers. *Issues in Science and Technology Librarianship, 67*.

Garner, J., Horwood, L., & Sullivan, S. (2001). The place of eprints in scholarly information delivery. *Online Information Review, 25*(4), 250-253.

Gentil-Beccot, A., Mele, S., & Brooks, T.C. (2009). Citing and reading behaviours in high-energy physics. How a community stopped worrying about journals and learned to love repositories. *arXiv:0906.5418*.

Ginsparg, P. (1994). First steps towards electronic research communication. *Computers in Physics, 8*(4), 390-296.

Ginsparg, P. (2008). The global village pioneers. *Learned Publishing, 21*, 95-100.

Ginsparg, P. (2011). arXiv at 20. *Nature, 476*, 146-147.

Goldschmidt-Clermont, L. (1965). Communication patterns in high-energy physics.

Haque, A.-u., & Ginsparg, P. (2009). Positional effects on citation and readership in arXiv. *Journal of the American Society for Information Science & Technology, 60*(11), 2203-2218.